%% file: paper.tex
\documentclass[twocolumn,showpacs,aps,prl,superscriptaddress]{revtex4}

\usepackage{graphicx}
\usepackage{dcolumn}
\usepackage{amsmath}
\usepackage{amssymb}
\usepackage{epsfig}
\usepackage{amsgen}
\usepackage{amsmath}
\usepackage{amsgen}
\usepackage{amsmath}
\usepackage{amsfonts}

\input pubboard/babarsym

\newcommand{\z}{\ensuremath{{\mathsf z}}\xspace}
\newcommand{\Imz}{\ensuremath{\rm Im\, \z}}
\newcommand{\Rez}{\ensuremath{\rm Re\, \z}}
\newcommand{\dG}{\ensuremath{ \Delta \Gamma }}
\newcommand{\absqop}{\ensuremath{|q/p|}}

\newcommand{\AT}{\ensuremath{A_{T/CP}}}
\newcommand{\ACPT}{\ensuremath{A_{CPT/CP}}}
\newcommand{\dGRez}{\ensuremath{\dG \times \Rez}}

\def\dt {\ensuremath{\Delta{\rm t}}}
\def\dm {\ensuremath{\Delta m}}

\newcommand{\BABARPubYear}    {06}
\newcommand{\BABARPubNumber}  {11}

\newcommand{\SLACPubNumber} {yy}

\def\figurebox#1#2#3{%
    \def\arg{#3}%
    \ifx\arg\empty
    {\hfill\vbox{\hsize#2\hrule\hbox to #2{\vrule\hfill\vbox to #1{\hsize#2\vfill}\vrule}\hrule}\hfill}%
    \else
    {\hfill\epsfbox{#3}\hfill}%
    \fi}

\begin{document}
\preprint{\babar-PUB-\BABARPubYear/\BABARPubNumber}
\preprint{SLAC-PUB-\SLACPubNumber}

\begin{flushleft}
\babar-PUB-\BABARPubYear/\BABARPubNumber\\
SLAC-PUB-11787 
\end{flushleft}
\vskip -1mm

\title{
{\large \bf \boldmath
Search for  \T, \CP and \CPT Violation in \Bz -\Bzb Mixing
with Inclusive Dilepton Events}
}


\input pubboard/authors_feb2006.tex

\date{\today}

\begin{abstract}
We report the results of a search for  \T, \CP and \CPT violation in \Bz -\Bzb mixing using an
inclusive dilepton sample collected by the \babar\ experiment at the
\pep2\ \BF. 
Using a sample of 232 million \BB\ pairs,
with a simultaneous likelihood fit of the same-sign
and opposite-sign dileptons, we measure the \T and \CP violation parameter 
$\absqop \rm  -1 =( -0.8 \pm 2.7(stat.)\pm 1.9(syst.))\times 10^{-3}$,
and the \CPT and \CP parameters 
$\Imz = (-13.9 \pm 7.3 (stat.)\pm 3.2(syst.))\times 10^{-3}$ 
and $\dGRez = (-7.1 \pm 3.9(stat.)\pm 2.0(syst.))\times 10^{-3}\ps^{-1}$.
The statistical correlation between the measurements of $\Imz$ and \dGRez\ is 76 \%. 
\end{abstract}


\maketitle


Since the first observation of \CP violation in 1964~\cite{CP64}, the neutral kaon system
has provided many results probing the \CPT and \T discrete
symmetries~\cite{CPLEAR99} in \Kz -\Kzb  mixing. Similarly,
 the \babar\ experiment can investigate \T,
 \CP, and \CPT violation in \Bz -\Bzb  mixing.

The physical states (solutions of the complex effective Hamiltonian 
for  the \Bz -\Bzb system)~\cite{BaBarCPT} can be written as
\begin{eqnarray}
|B_L\rangle&=&p\sqrt{1-\z}|\Bz\rangle +q\sqrt{1+\z}|\Bzb\rangle,\nonumber \\
|B_H\rangle&=&p\sqrt{1+\z}|\Bz\rangle -q\sqrt{1-\z}|\Bzb\rangle.\nonumber
\end{eqnarray}
where $H$ and $L$ stand for Heavy and Light.
In the case of \CPT invariance, the complex parameter \z is equal to 0. Similarly, \T
invariance leads to $\absqop=1$. Finally, \CP invariance requires 
both $\absqop=1$ and $\z~=~0$. 

Inclusive dilepton events, where both $B$ mesons decay
semileptonically   $b \to X\ell\nu$ ($l=e$ or $\mu$),
 represent 4\% of all \upsbb
decays and  provide a very large sample  to study \T, \CPT 
and \CP violation in mixing.
In the direct $b \to \ell$  decay process,
the flavor $\Bz(\Bzb)$  is tagged by the charge of the lepton $\ellp(\ellm)$.

At the \FourS resonance, neutral $B$ mesons are produced in a coherent p-wave state.
At the instant that the first B meson decays, the second B meson has the 
opposite flavor.  Then, the second B meson will continue to evolve in 
time.
Defining the time difference as $\dt = t^{+}-t^{-}$ where $t^{+}(t^{-})$ is 
the decay time of the neutral $B$ tagged by $\ellp(\ellm)$, 
and neglecting second order terms in \z,
the decay
rates for the three configurations ($\ellp\ellp$, $\ellm\ellm$
and $\ellp\ellm$) are given by \\
\vskip -.5cm
{\small
\begin{eqnarray}
N^{++} &\propto & \frac{e^{-\Gamma |\dt|}}{2}  |\frac{p}{q}|^2 
\Bigl\{ \cosh(\frac{\dG \dt}{2})
  -\cos(\dm \dt) \Bigr\},\nonumber \\
N^{--} &\propto &  \frac{e^{-\Gamma|\dt|}}{2} |\frac{q}{p}|^2  
\Bigl\{ \cosh(\frac{\dG \dt}{2})
  - \cos(\dm \dt) \Bigr\}, \nonumber\\
 N^{+-}&\propto & \frac{e^{-\Gamma |\dt|}}{2}
\Bigl\{ \cosh(\frac{\dG\! \dt}{2}) \nonumber 
  - 2\,\Rez\, \sinh(\frac{\dG\! \dt}{2}) \nonumber \\
 & &   \qquad \quad \;  + \cos(\dm\! \dt) + 2\,\Imz\,\sin(\dm\! \dt)\Bigr\},\label{eq:decayrate}
\end{eqnarray}
}
\hskip -.18cm
where $\dm$ is the \Bz-\Bzb\ oscillation frequency,  
$\Gamma$ is the average neutral $B$ decay rate and \dG\ is the decay rate difference between
the two physical states. 

The same-sign dilepton asymmetry $\AT$, between the two oscillation probabilities
$P(\Bzb \to \Bz)$ and $P(\Bz \to \Bzb)$  probes both \T and \CP symmetries
and can be expressed in terms of \absqop:
\begin{eqnarray}
\AT & =&  \frac {P(\Bzb \to \Bz)-P(\Bz \to \Bzb)}
                       {P(\Bzb \to \Bz)+P(\Bz \to \Bzb)}\nonumber \\
&=&  \frac {N^{++} - N^{--}}{N^{++} + N^{--}} 
=  \frac {1-|q/p|^4}{1+|q/p|^4}.
\label{eq:at}
\end{eqnarray}
Standard Model calculations~\cite{SM} predict
the size of this asymmetry to be at or below $10^{-3}$. 
A large measured value would be an
indication of new physics.

Similarly, the opposite-sign dilepton asymmetry, $\ACPT$, between events with
$\dt >0$ and $\dt<0$ compares the  $\Bz \to \Bz$ and
$\Bzb \to \Bzb$  probabilities and is sensitive to  \CPT and \CP violation.
 This asymmetry is given by
\vskip -.5cm
{\small
\begin{eqnarray}
\ACPT(|\dt|)& =&  \frac {P(\Bz \to \Bz)-P(\Bzb \to \Bzb)}
                       {P(\Bz \to \Bz) + P(\Bzb \to \Bzb)}
		       \nonumber\\
&=&    \frac {N^{+-}(\dt>0) - N^{+-}(\dt<0)}{N^{+-}(\dt>0) + N^{+-}(\dt<0)}
\nonumber\\ 
& \simeq & 2\frac{\Imz\sin(\dm\!\dt)\! - \!\Rez\sinh(\frac{\dG\! \dt}{2})}
{\cosh(\frac{\dG\! \dt}{2}) + \cos(\dm\!\dt)}.
\label{eq:acpt}
\end{eqnarray}
}
As $|\dG|/\Gamma \ll 1$~\cite{BaBarCPT}, we have $\Rez\sinh(\dG\dt/2) \simeq  \dGRez \times (\dt/2) $ and
this asymmetry is  not sensitive to  the $C\!PT$-violating term \Rez\ alone,
but to the product \dGRez. 

In this Letter, we present  measurements of  \absqop,  \Imz\ and \dGRez\
with a simultaneous likelihood fit of the same-sign and opposite-sign dilepton 
$\dt$ distributions. In the $\cosh(\dG \dt / 2)$ term,
we fix $|\dG|$ to $0.005\ps^{-1}$, the value reported in~\cite{BaBarCPT}
with a 90\% confidence-level limit of $0.055\ps^{-1}$.

This study is performed  with events collected by  the
\babar\ detector~\cite{BaBarNIM01} 
at the \pep2  asymmetric-energy \BF\ between October 1999 and July 2004.
The integrated luminosity of this sample is about 211 \invfb recorded
at the \FourS resonance (``on-resonance'') (232 million \BB\ pairs)
and about 16 \invfb recorded about 40 \mev below the \FourS\  resonance (``off-resonance'').


The event selection is identical to that described
in~\cite{BaBarAT}. 
Non-\BB events are suppressed by
applying requirements on the  ratio of second to zeroth order Fox-Wolfram moments~\cite{FW},
 the squared invariant mass, the aplanarity and the number of charged tracks
of the event.

Lepton candidate tracks must  have at
least 12 hits in the drift chamber, at least one $z$-coordinate
hit in the silicon vertex tracker (SVT),
and  a momentum in the \FourS center-of-mass system between
0.8 and 2.3 \gevc.
Electrons are selected by  requirements
on the ratio of the energy deposited in the electromagnetic calorimeter
and the momentum measured in the drift chamber.
Muons are identified through the energy released in the calorimeter, as
well as the strip multiplicity, track continuity, and penetration depth in the
instrumented flux return. Lepton candidates are rejected if their signal in 
the detector of internally reflected Cherenkov light is consistent with that of 
a kaon or a proton. The electron and muon selection
efficiencies are about 85\% and 55\%, with pion misidentification probabilities around
0.2\% and 3\%, respectively.

Electrons from photon conversions are identified and rejected
with a negligible loss of efficiency for signal events.
Leptons from \jpsi\ and $\psi (2S)$ decays are identified by pairing
them with other oppositely-charged candidates of the same lepton species,
selected with looser criteria. 

The separation between {\it direct} leptons $(b \to \ell)$ and background from 
the $b\rightarrow c\rightarrow \ell$ decay chain ({\it{cascade decays}}) 
is achieved with a neural network that combines five discriminating variables: 
the momenta and opening angles of the two lepton candidates,
and the total visible energy and missing momentum of the event, all
computed in the \FourS\ rest frame. 
 
Of the original sample of 232 million \BB\ pairs, 
1.4 million pass this dilepton selection.


Since the asymmetry $\AT$ is expected to be small, we have
determined the possible charge asymmetries induced by the detection and
reconstruction of electrons and muons. The charge asymmetries
are defined by 
$ a \equiv (\varepsilon^+ - \varepsilon^-)/(\varepsilon^+ + \varepsilon^-)$
where $\varepsilon^+ (\varepsilon^-)$ is the efficiency for 
positive and negative particles. As the lepton
efficiencies and  purities
depend on their allowed phase space, we consider 
separately the asymmetry for the higher and lower momentum lepton,
respectively, $a_1$ and $a_2$. 

The charge asymmetry of track reconstruction is measured in the data
by comparing tracks reconstructed using only the SVT with those passing 
the dilepton track selection,  
obtaining $a_{trk}=(0.8\pm0.2)\times 10^{-3}$.

The lepton identification efficiencies are measured
as a function of total momentum and  polar and azimuthal angles,
with  a control sample of 
radiative Bhabha events for  electrons, and with a
 $ee \to \mu\mu\gamma$ control sample  for  muons.
The  misidentification probabilities are determined with control
samples of kaons produced in $D^{*+}\to\pi^+D^0\to\pi^+K^-\pi^+$ 
(and charge conjugate) decays, pions produced in $K_S\to\pipi$ decays,
one-prong and three-prong $\tau$ decays, and protons
produced in $\Lambda$ decays.

The control samples show that the muon track reconstruction
efficiency has a charge asymmetry reaching $\sim 5\times10^{-3}$ and that the positive
kaons are more likely than negative kaons to be misidentified as
muons at the 20-30\% level.
As a consequence, in the likelihood fit (described below), we float the
charge asymmetries $a_{\mu}^{dir}$ and $a_{\mu}^{casc}$ for direct and cascade muons.

For electrons, the charge asymmetry averaged over the signal phase space 
is $a_{e}=(0.4\pm0.2)\times 10^{-3}$ 
and we find that antiprotons with momentum  $\sim 1 \gevc$
are significantly more likely than protons to be misidentified,
due to annihilation with nucleons in the calorimeter material.
Based on the charge asymmetry in tracking and in identification, we fix
the charge asymmetry for the direct electrons with the higher momentum to 
$a_{e_1}^{dir}=1.2 \times 10^{-3}$. For the lower momentum direct electrons and the cascade
 electrons, for which antiprotons contamination is more important, we correct 
the initial charge asymmetry by the fraction of
antiprotons estimated with  generic \BB\ Monte Carlo samples and the proton control sample,
this gives the following charge asymmetries: $a_{e_2}^{dir}=0.8 \times 10^{-3}$,
$a_{e_1}^{casc}=0.5 \times 10^{-3}$, and $a_{e_2}^{casc}=0.2 \times 10^{-3}$.

In the inclusive approach used here, the $z$ coordinate of the $B$ decay point 
is approximated by 
the $z$ position of the point of closest approach between the lepton
candidate and an estimate of the \FourS decay point in the transverse plane. The
\FourS decay point is obtained by fitting the two lepton tracks to a common vertex
in the transverse plane that is constrained to be consistent with  the beam-spot 
position. The proper time difference $\dt$ between the two $B$ meson decays
is determined from  $\Delta z = z^+ - z^-$, the difference in $z$ between the leptons 
 $\ell^+$ and $\ell^-$, by $\dt=\Delta z/ \langle\beta\gamma\rangle c$ with a nominal Lorentz boost  
$\langle \beta \gamma \rangle = 0.55$. In case of same-sign dileptons, the sign of
\dt\ is chosen randomly.

We model the contributions to our sample from \BB\  decays using
five categories of events, $i$, each represented by a probability density
function (PDF) in $\dt$,  ${\cal P}_{i}^{n,c}$. Their shapes  are determined using 
the $\BzBzb$ $(n)$ and $\BpBm$ $(c)$ Monte Carlo simulation separately, 
 with the approach described in~\cite{BaBardm}.

The five categories are the following. 
First, the pure signal events  with two direct leptons ($sig$)  represent 81\% of the 
\BB\  events and give information on the \T, \CPT 
and \CP\ parameters.
Then, we consider two categories of cascade decays: those with a
direct lepton and cascade lepton from the opposite $B$ decays
($obc$), and those with direct lepton and cascade lepton from the same $B$ decay
($sbc$). According to generic \BB\ Monte Carlo simulation, their contributions are around 
9\% and  4\% respectively. In addition, 
3\% of the dilepton events  originate from the decay chain $b \to \tau^- \to \ell^-$ ($1d1\tau$)
which tags the $B$ flavor correctly. Finally, the remaining events ($other$) consist mainly of
one direct lepton and one lepton from charmonium resonances in the $B$
decays.

The $sig$ event PDF,  ${\cal P}_{sig}^{n,c}$, 
is obtained by the convolution of an oscillatory term 
containing the \T, \CPT 
and \CP\ parameters (Eq.~\ref{eq:decayrate}) for neutral $B$ decays 
or an exponential function for charged $B$ decays, 
with a resolution function which is the sum of three Gaussians. 
The widths of the core and tail  
Gaussians and the fractions of core and outlier Gaussians are
free parameters in the fit. The width of the outlier Gaussian 
is fixed to $8\ps$. The means of the Gaussians are fixed
to zero \cite{bib:BABAR-s2b}.

The $obc$ event PDF,  ${\cal P}_{obc}^{n,c}$, is modeled by the convolution
of $\dt$-dependent terms of a form similar to those of  the signal with 
a resolution function which takes into account the effect of the charmed meson lifetimes. 
Since both short-lived $D^0$ and $D_S$, and long-lived $D^+$ mesons
are involved in cascade decays, the resolution function for the long-lived and short-lived  components 
is a sum of three Gaussians, which are convoluted with  double-sided exponentials. 
To correct the effect of possible outliers not observed with the Monte Carlo simulation, the
fraction of the third Gaussian is free in the fit. Similarly, we take the effect of the 
charmed mesons into account in the $sbc$ event PDF, ${\cal P}_{sbc}^{n,c}$.

The PDF for $1d1\tau$ events, ${\cal P}_{1d1\tau}^{n,c}$ is similar to that of the $sig$ events. 
The resolution function used takes into account the $\tau$ lifetime effect and is chosen 
to be two Gaussians convoluted with two double-sided exponentials. 
Finally, the PDF for the remaining events, 
${\cal P}_{other}^{n,c}$, is the convolution  of 
an exponential function with an effective lifetime and two Gaussians.  

The fractions ($f_{sbc}^{n,c}$,$f_{1d1\tau}^{n,c}$ and
 $f_{other}^{n,c}$) of $sbc$, $1d1\tau$ and $other$ events,  
are determined directly with the $\BzBzb$ and $\BpBm$ Monte Carlo simulation. The fractions
$f_{obc}^{n,c}$ of $obc$ events  are fitted to the data, 
constraining the ratio $f_{obc}^{n}/f_{obc}^{c}$ to the estimate
obtained with Monte Carlo samples. The fraction $f_{+-}$ of  $\BpBm$ events is
determined from the data themselves.

The last component of the dilepton sample originates from non-\BB\ events,
mainly continuum events,
and has been estimated using off-resonance events to represent a fraction $f_{cont}= (3.1\pm0.1)\%$ 
of the data set. 
To model its PDF we use off-resonance events with looser cuts 
and  on-resonance events that fail the 
continuum-rejection cut on the Fox-Wolfram moment ratio.
The charge asymmetries $a^{cont}_{e,\mu}$ obtained with the two samples
are consistent with zero at the 1\% level and thus are fixed to zero in the likelihood.

The $\T/\CP$ and $\CPT/\CP$ violation parameters are extracted from a
binned maximum likelihood fit of the events that pass the dilepton selection. 
The likelihood ${\cal L}$ combines the charge asymmetries in detection and the time-dependent PDFs 
described previously. As the charge asymmetries are significantly
different for electrons and muons, we split the sample into four lepton combinations: $ee$, $e\mu$, $\mu e$ and
$\mu\mu$, in which the first lepton has the higher momentum.

The likelihood is given by
{\small
\begin{eqnarray}
  { \cal L}(\dt)   & = & (1+q_1a_{{\rm f}_1}^{cont})(1+q_2a_{{\rm f}_2}^{cont})f_{cont} {\cal P}_{cont} \nonumber \\
  && + (1-f_{cont}) \{  f_{+-}{\cal P}_{B^+B^-}
  + (1-f_{+-}){\cal P}_{B^0\overline B^0} \} \nonumber \\
  {\cal P}_{\BzBzb} & = & (1-f^n_{sig})(1+q_1a_{{\rm f}_1}^{casc})(1+q_2a_{{\rm f}_2}^{casc}) {\cal P}^n_{casc} \nonumber \\
  && + f^n_{sig}(1+q_1a_{{\rm f}_1}^{dir})(1+q_2a_{{\rm f}_2}^{dir}) {\cal P}^n_{sig} \nonumber\\
  {\cal P}_{\BpBm} & = & (1-f^c_{sig})(1+q_1a_{{\rm f}_1}^{casc})(1+q_2a_{{\rm f}_2}^{casc}) {\cal P}^c_{casc}\nonumber \\
  && + f^c_{sig}(1+q_1a_{{\rm f}_1}^{dir})(1+q_2a_{{\rm f}_2}^{dir}) {\cal P}^c_{sig} \nonumber \\
  {\cal P}^{n,c}_{casc} & = & f^{n,c}_{other} {\cal P}^{n,c}_{other} + f^{n,c}_{1d1\tau} {\cal P}^{n,c}_{1d1\tau} 
  + f^{n,c}_{sbc} {\cal P}^{n,c}_{sbc} + f^{n,c}_{obc} {\cal P}^{n,c}_{obc}, \nonumber
\end{eqnarray}
}
where  $q_1$, $q_2$, ${\rm f}_1$ and ${\rm f}_2$   are the charges and the flavors $(e,\mu)$
of the two leptons.  

\begin{figure}[hbtp]
\begin{center}
\mbox{\epsfig{file=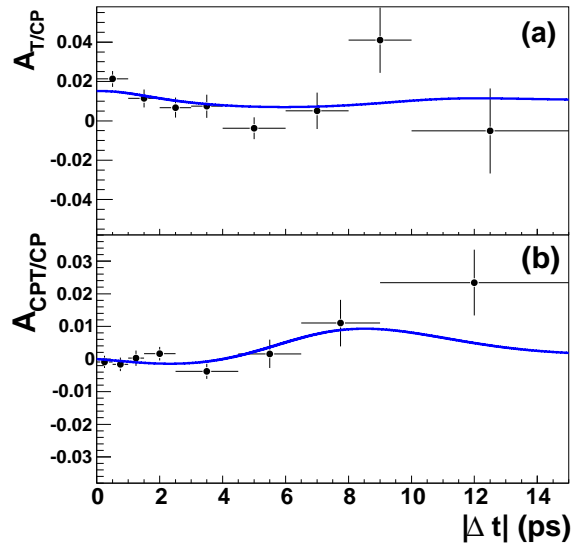,height=7.5cm}}
\end{center}
\caption{ (a) $\AT$ asymmetry between $(\ell^+,\ell^+)$ and $(\ell^-,\ell^-)$.
A larger charge asymmetry for cascade muons, dominant at small $|\dt|$, explains the non-flatness
of the curve. 
 (b)  $\ACPT$ asymmetry between  $(\ell^+,\ell^-)$ 
dileptons with $\dt>0$ and $\dt<0$.}
\label{fig:AsyData}
\end{figure}

The likelihood fit gives $\absqop  -1 =( -0.8 \pm 2.7)\times 10^{-3}$, 
$\Imz = (-13.9 \pm 7.3 )\times 10^{-3}$, 
and $\dGRez = (-7.1 \pm 3.9)\times 10^{-3}\ps^{-1}$.
The correlation between the measurements of $\Imz$ and \dGRez\ is 76 \%.
If we fix $\Delta \Gamma = 0$, we obtain $\Imz = (-3.7 \pm 4.6 )\times 10^{-3}$.
Figure~\ref{fig:AsyData} shows  the
 $\AT$ asymmetry between $(\ell^+,\ell^+)$ and $(\ell^-,\ell^-)$
dileptons defined in Eq.~\ref{eq:at} and the $\ACPT$ asymmetry between  $(\ell^+,\ell^-)$ 
dileptons with $\dt>0$ and $\dt<0$ defined in Eq.~\ref{eq:acpt}.

There are several sources of systematic uncertainty in these measurements.  
To determine their effect,
we vary each source of systematic uncertainty by its known or estimated 
uncertainty, and take the resulting deviation in the \CP\ parameter 
as its systematic uncertainty.

For \absqop, the most important systematic errors are due to uncertainties on
electron charge asymmetries. A 
 $1.4\times 10^{-3}$ deviation of \absqop\ is observed  by shifting simultaneously 
the  electron charge asymmetries by $1.0\times 10^{-3}$ which corresponds
to the uncertainty estimated with Monte Carlo and control samples. 
The systematic uncertainty related to 
the charge asymmetry due to the tracking is estimated by  randomly removing a fraction
equal to $1.6 \times 10^{-3}$ of the negative tracks from our data sample. 
This fraction has been determined from an independent data control sample.  A 
$1.0\times 10^{-3}$ deviation of \absqop\ is observed. Similarly,  the
$1\%$ uncertainty on charge asymmetry for non-\BB\ backgrounds induces a
systematic error of $0.6\times 10^{-3}$.

The widths of the first and second Gaussian of the resolution function for the $obc$ and $sbc$ 
categories as well as the pseudo-lifetime for the $1d1\tau$ and $other$ 
categories are varied separately by 10\%. This variation is motivated by the comparison of the 
fitted parameters of the signal resolution function obtained on generic \BB\ Monte Carlo samples
and on data being in agreement at 10\% level.
The fractions of the short-lived and long-lived 
charmed meson components  for $obc$ and $sbc$ are varied by 10 \%. 

We have also varied the parameters  $\dm$, $\tau_{\Bz}$ and $\tau_{\Bpm}$ independently 
within their known uncertainties~\cite{bib:HFAG} and \dG\ from $10^{-5}$ to $ 0.1$.  
Finally, one of the dominant systematic 
errors on \dGRez\ is imperfect knowledge of the 
absolute $z$ scale of the detector and the residual uncertainties in the 
SVT local alignment, giving an error of $1.2 \times 10^{-3}\ps^{-1}$.

\begin{table} [htbp]
\begin{center}
\caption{Summary of systematic errors for \absqop, \Imz, and \dGRez\ measurements.}
\begin{tabular}{lccc}
\hline
\hline
{\bf Systematic Effects}  &  $\sigma(\absqop)$ & $\sigma(\Imz)$ 
& $\sigma(\dG\!\! \times\! \Rez)$\\
 & $(\times 10^{-3})$ &$ (\times 10^{-3})$ & $(\times 10^{-3}\ps^{-1})$\\
\hline
Ch. asym. of non-\BB\ bkg & $ 0.6$ & $ 0.0$ 	& $ 0.0$ \\
Ch. asym. in tracking 		& $ 1.0$ & $ 0.0$ 	& $ 0.0$\\
Ch. asym. of electrons 		& $ 1.4$ & $ 0.0$ 	& $ 0.0$\\
PDF modeling 				& $ 0.3$ & $ 2.5$ 	& $ 1.2$\\
Fraction of bkg components 	& $ 0.2$ & $ 0.4$ 	& $ 0.1$\\
$\dm$, $\tau_{\Bz}$, $\tau_{\Bpm}$ and $\Delta \Gamma$  & $ 0.2$ & $ 1.9$ 	& $ 1.1$\\
SVT alignment 				& $ 0.5$ & $ 0.6$ 	& $ 1.2$\\
\hline
Total 					& $ 1.9$  & $ 3.2$	& $ 2.0$\\
\hline
\hline
\end{tabular}
\label{tab:Syst}
\end{center}
\end{table} 

For each parameter, the total 
systematic error is the sum in quadrature of the estimated systematic 
errors from each source, as summarized in Table~\ref{tab:Syst}.
When we assume $\Delta \Gamma = 0$,
the systematic error for \Imz\ is $  2.9\times 10^{-3}$.

If we compare our results to  $\dGRez= 0.0$ and $\Imz=0.0$ (no \CPT\ 
violation case), the $\chi^2$ is 3.25 for 2 degrees of freedom, which gives a 
confidence level  of 19.7\%. 
Finally, assuming $\Delta \Gamma = 0$, we obtain $\Imz = (-3.7 \pm 4.6(stat.)\pm 2.9(syst.) )\times 10^{-3}$.


In summary with the 1999-2004 data  ($232\times 10^{6}$ \BB\ pairs), we 
have performed a simultaneous likelihood fit of the same-sign and
opposite-sign dileptons. We  measure the independent parameters
governing  \CP and \T violation, 
and the \CPT and \CP violation parameters. 
The results are
\begin{eqnarray*}
\absqop -1 & = & (-0.8 \pm 2.7{\rm (stat.)}  \pm 1.9 {\rm(syst.)})\times 10^{-3},\\
\Imz  & = & (-13.9  \pm 7.3 {\rm (stat.)} \pm 3.2{\rm(syst.)})\times 10^{-3},\\
\dGRez & = & (-7.1 \pm 3.9 {\rm (stat.)} \pm 2.0{\rm(syst.)})\times 10^{-3}\ps^{-1}.
\end{eqnarray*}
These measurements are a clear improvement over the most
precise results previously published~\cite{BaBarCPT,BelleAT}.
The new measurement of \absqop\ is consistent with
the Standard Model predictions~\cite{SM}.

\input pubboard/acknow_PRL

\end{document}

%% file: pubboard/authors_feb2006.tex
%
\author{B.~Aubert}
\author{R.~Barate}
\author{M.~Bona}
\author{D.~Boutigny}
\author{F.~Couderc}
\author{Y.~Karyotakis}
\author{J.~P.~Lees}
\author{V.~Poireau}
\author{V.~Tisserand}
\author{A.~Zghiche}
\affiliation{Laboratoire de Physique des Particules, F-74941 Annecy-le-Vieux, France }
% ...
\collaboration{The \babar\ Collaboration}
\noaffiliation

%% file: pubboard/acknow_PRL.tex
We are grateful for the excellent luminosity and machine conditions
provided by our \pep2\ colleagues, 
and for the substantial dedicated effort from
the computing organizations that support \babar.
The collaborating institutions wish to thank 
SLAC for its support and kind hospitality. 
This work is supported by
DOE
and NSF (USA),
NSERC (Canada),
IHEP (China),
CEA and
CNRS-IN2P3
(France),
BMBF and DFG
(Germany),
INFN (Italy),
FOM (The Netherlands),
NFR (Norway),
MIST (Russia), and
PPARC (United Kingdom). 
Individuals have received support from CONACyT (Mexico), 
Marie Curie EIF (European Union),
the A.~P.~Sloan Foundation, 
the Research Corporation,
and the Alexander von Humboldt Foundation.